\documentclass[twoside,prd,superscriptaddress,showpacs,showkeys,twocolumn,nofootinbib]{revtex4}

\usepackage[active]{srcltx}

\usepackage{amsmath}
\usepackage{slashed}
\usepackage{graphicx}
\usepackage{hyperref}
\usepackage{wasysym} 

\newcommand{\bra}[1]{\langle#1\vert}
\newcommand{\ket}[1]{\vert#1\rangle}
\newcommand{\I}{\ensuremath{\mathrm{i}}}
\renewcommand{\Re}{\mathop{\rm Re}\nolimits}
\renewcommand{\Im}{\mathop{\rm Im}\nolimits}
\newcommand{\eL}{\mathcal{L}}
\renewcommand{\d}{\ensuremath{\mathrm{d}}}
\newcommand{\hc}{\ensuremath{\mathrm{h.c.}}}
\newcommand{\BR}{\mathop{\rm BR}\nolimits}
\newcommand{\heaviside}{\mathop{\theta}\nolimits}

\usepackage{color}
\usepackage{ulem} 
\definecolor{myred}{rgb}{1,0,0}
\definecolor{mygreen}{rgb}{0,0.8,0.2}
\definecolor{myblue}{rgb}{0,0,1}

\begin{document}
\title{Dynamical electroweak symmetry breaking due to strong Yukawa interactions}

\author{Petr Bene\v s}
\email{benes@ujf.cas.cz}
\affiliation{Department of Theoretical Physics, Nuclear Physics Institute ASCR, \v Re\v z, Czech Republic}
\affiliation{Faculty of Mathematics and Physics, Charles University, Prague, Czech Republic}

\author{Tom\'a\v s Brauner}
\email{brauner@ujf.cas.cz}
\affiliation{Department of Theoretical Physics, Nuclear Physics Institute ASCR, \v Re\v z, Czech Republic}
\affiliation{Institut f\"{u}r Theoretische Physik, Goethe-Universit\"{a}t, Frankfurt am Main, Germany}

\author{Adam Smetana}
\email{adam.smetana@ujf.cas.cz}
\affiliation{Department of Theoretical Physics, Nuclear Physics Institute ASCR, \v Re\v z, Czech Republic}
\affiliation{Faculty of Mathematics and Physics, Charles University, Prague, Czech Republic}

\begin{abstract}
We present a new mechanism for electroweak symmetry breaking (EWSB) based on a strong Yukawa dynamics. We consider an $SU(2)_L \times U(1)_Y$ gauge invariant model endowed with the usual Standard model fermion multiplets and with two massive scalar doublets. We show that, unlike in the Standard model, EWSB is possible even with vanishing vacuum expectation values of the scalars. Such EWSB is achieved dynamically by means of the (presumably strong) Yukawa couplings and manifests itself by the emergence of fermion and gauge boson masses and scalar mass-splittings, which are expressed in a closed form in terms of the fermion and scalar proper self-energies. The `would-be' Nambu--Goldstone bosons are shown to be composites of both the fermions and the scalars. We demonstrate that the simplest version of the model is compatible with basic experimental constraints.
\end{abstract}

\maketitle

\section{Introduction}
The majority of current models of electroweak symmetry breaking (EWSB) can
be divided into two large classes, according to the mechanism the symmetry
breaking is achieved (omitting exotic scenarios such as extra dimensions): (i)
The electroweak symmetry is broken at tree level by non-vanishing vacuum
expectation value(s) (VEV(s)) of some elementary scalar field(s). This class
includes the Standard model (SM) \cite{Weinberg:1967tq} and its various
extensions like the two-Higgs-doublet models \cite{Lee:1973iz}, the minimal
supersymmetric Standard model \cite{Haber:1984rc}, or the little Higgs
models \cite{ArkaniHamed:2001nc,Arkani-Hamed:2002qx}. (ii) The electroweak
symmetry is broken non-perturbatively by strong dynamics of some new, yet
unobserved, interactions. This class includes theories with new strong gauge
interactions such as (extended) technicolor
\cite{Weinberg:1979bn,Susskind:1978ms,Dimopoulos:1979es,Eichten:1979ah}, top
condensate \cite{Miransky:1988xi,Bardeen:1989ds}, or
topcolor \cite{Hill:1991at,Hill:1994hp}, as well as effective models of the
Nambu--Jona-Lasinio type \cite{Nambu:1961tp,Hosek:1982cz}.

In this paper we study a different scenario, which combines features of both
classes mentioned above (elementary scalars and strong dynamics): The electroweak
symmetry is broken non-perturbatively by strong Yukawa interactions (a similar
approach can be found in \cite{Wetterich:2006ii,Luty:2000fj} and, in the context
of non-relativistic systems, in
\cite{Imry,PhysRevLett.89.075301,Alexandrov2004}). We believe that this option
has certain advantages. For example, it allows to distinguish otherwise
identical fermions at the level of Lagrangian by different Yukawa couplings and
hence avoid dangerous (because unobserved) Nambu--Goldstone (NG) bosons
resulting from the spontaneous breaking of the large flavor global symmetries.
The idea has already been studied in our previous paper \cite{Benes:2006ny}
using a prototype model with Abelian gauge symmetry. We have shown that the
dynamically generated masses of different fermions may differ by orders of
magnitude even with Yukawa coupling ratios of order one. We thus suppose that
the present mechanism could shed some new light on the problem of the fermion
mass hierarchy in SM.

Encouraged by the previous model study, we investigate in this paper the
possibility of EWSB by the same mechanism, considering for simplicity a
single family of SM fermions. At this exploratory stage we do not want to
make any kind of phenomenological predictions, our aim is merely to demonstrate that the
extension of the scenario to the electroweak symmetry brings no new principial problems.
In Sec.~\ref{sec_model}, we first introduce a model with global $SU(2)_L\times
U(1)_Y$ symmetry, which is broken dynamically by the strong Yukawa couplings. In Sec.~\ref{sec_gauged}, we then gauge the symmetry, which induces dynamical EWSB.
In Sec.~\ref{sec_results} we present numerical results and perform a
basic check of consistency with precision electroweak measurements.
The outlook for future work as well as potential applications in other fields of
physics are discussed in the conclusions.

\section{The model}
\label{sec_model}
We consider an $SU(2)_L\times U(1)_Y$ invariant Lagrangian, equipped with
one generation of the SM fermion multiplets: The left-handed
isospin doublets, denoted as $\ell_{L}=(\nu_L,e_L)^T$,
$q_{L}=(u_L,d_L)^T$, and the right-handed singlets $e_{R}$, $u_{R}$, $d_{R}$. In
addition, we introduce the right-handed neutrino singlet $\nu_{R}$ in order to
generate the neutrino mass. However, we will consider only the Dirac mass here,
since the Majorana mass would require a more elaborate formalism. We also
introduce two complex scalar doublets, $S=(S^{(+)},S^{(0)})^T$ and
$N=(N^{(0)},N^{(-)})^T$. Unlike in SM, they have ordinary masses $M_S$, $M_N$
(i.e., $M_{S,N}^2>0$), which implies they do not condense at tree level. Hence
there is no need for the scalar self-couplings to stabilize the vacuum and they
will for simplicity be neglected.

Of key importance are the Yukawa couplings of the fermions to $S$, $N$, by assumption
responsible for EWSB:
\begin{eqnarray}
\eL_{\mathrm{Yukawa}} &=& y_{e} \bar \ell_{L}  e_{R} S +  y_{\nu}\bar \ell_{L}  \nu_{R} N+
y_{d}\bar q_{L} d_{R} S + y_{u}\bar q_{L}  u_{R} N \nonumber
\\
&& {}+ \hc
\label{L_Y}
\end{eqnarray}
The Yukawa couplings can always be made
real by a suitable change of phases of the felds. For
simplicity we do not consider interactions of the charge conjugated scalar
doublets, $\tilde S \equiv \I \sigma_2 S^*$ and $\tilde N \equiv \I \sigma_2 N^*$.
Our `minimal' Yukawa interaction Lagrangian in fact turns out to be sufficient for
generating fermion and gauge boson masses.

In the present paper we deliberately neglect renormalization and running of
the Yukawa couplings. Within our model this is decently justified by the fact
that thanks to the structure of the interaction Lagrangian \eqref{L_Y}, the
one-loop \emph{perturbative} corrections to the Yukawa couplings are absent.
Throughout the calculation, the couplings are therefore implicitly assumed to be
fixed and renormalized at the scale of the scalar masses $M_{S,N}$.

Of course, once the couplings become large and moreover the chiral symmetry is
spontaneously broken, this na\"{\i}ve argument breaks down. We do not attempt to
determine the full nonperturbative flow of the couplings but merely note that
it can be analyzed using the renormalization-group techniques, as was done for
similar models in Refs.~\cite{Bornholdt:1992za,Gies:2003dp}. Our results, in
particular the existence of a critical coupling for spontaneous symmetry
breaking, are in qualitative agreement.

With the above remarks in mind, we assume the full fermion propagators
to have the form ($f=\nu,e,u,d$)
\begin{eqnarray}
\langle f \bar f \rangle = \I \left( \slashed{p}-\Sigma_{f,1}-\I\gamma_5\Sigma_{f,2}
\right) ^{-1}
\,,
\label{ansatz_S}
\end{eqnarray}
where the $\Sigma$'s are \emph{real} functions of $p^2$ (the argument will be often
omitted throughout the paper). Notice that the $\Sigma$'s are scalar functions, which
means that we also neglect the perturbative wave function renormalization.

Likewise, for the full scalar propagators we adopt the Ansatz ($X=S,N$)
\begin{subequations}
\label{ansatz_D}
\begin{eqnarray}
\langle \Phi_X^{\vphantom{\dag}}\Phi_X^\dag \rangle &=&
\I\left( \begin{array}{cc}
p^2 - M_X^2 & -\Pi_X \\
-\Pi_X^{*} & p^2 - M_X^2
\end{array} \right)^{-1}
\,,
\\
\langle \Phi_{SN}^{\vphantom{\dag}}\Phi_{SN}^\dag \rangle &=&
\I\left( \begin{array}{cc}
p^2 - M_S^2 & -\Pi_{SN} \\
-\Pi_{SN}^{*} & p^2 - M_N^2
\end{array} \right)^{-1}
\,,
\end{eqnarray}
\end{subequations}
written using the Nambu doublet notation $\Phi_X = (
X^{(0)}, X^{(0)\dag})^T$, $\Phi_{SN} = ( S^{(+)}, N^{(-)\dag})^T$. The $\Pi$'s are
\emph{complex} functions of $p^2$.

The spectrum can now be revealed by looking for the poles of the full
propagators. In the fermionic case, provided there are non-vanishing
self-energies
$\Sigma_{f} \equiv \Sigma_{f,1}+\I\Sigma_{f,2}$, one arrives at the simple pole equations
$p^2 - |\Sigma_f(p^2)|^2 = 0$. For the neutral scalar doublets $\Phi_X$ the pole equation
reads $(p^2 - M_X^2)^2 - |\Pi_X(p^2)|^2 = 0$. Its two solutions $M_{X1,2}$ are to
be interpreted as that instead of one electrically neutral, yet \emph{complex} scalar
field $X^{(0)}$ with the mass $M_X$ there are \emph{two real} scalar fields
$X^{(0)}_{1,2}$ with distinct masses $M_{X1,2}$ in the spectrum. These mass eigenstates
are apparently linear combinations of $\Re X^{(0)}$ and $\Im X^{(0)}$.
Similarly considering the charged doublet $\Phi_{SN}$, its pole equation is $(p^2 -
M_S^2)(p^2 - M_N^2) - |\Pi_{SN}(p^2)|^2 = 0$. In this case the two solutions $M_{SN1,2}$
indicate that the charged fields $S^{(+)}$ and $N^{(-)\dag}$ need to be mixed in order to
get the mass eigenstates. In principle, charge conservation also allows mixing of
$S^{(0)}$ and $N^{(0)}$, but this only appears at the two-loop level.

\begin{figure}[t]
\begin{center}
\includegraphics[width=0.4\textwidth]{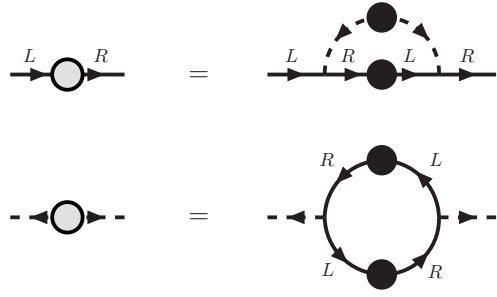}
\caption[]{Schematic diagrammatic representation of the
Schwinger--Dyson equations \eqref{eqn:SD}. Full blobs represent the full
propagators, the shaded blobs the one-particle-irreducible propagators
proportional to the $\Sigma$'s and $\Pi$'s, respectively. The letters
$L,R$ denote the fermion chiralities.}
\label{fig_SD}
\end{center}
\end{figure}

For later purposes it is convenient to define the `anomalous' propagators,
\begin{subequations}
\begin{eqnarray}
S_f(p) &=& \frac{\Sigma_{f}}{p^2-|\Sigma_{f}|^2}
\,,
\\
D_X(p) &=& \frac{\Pi_X}{(p^2-M_X^2)^2-|\Pi_X|^2}
\,,
\\
D_{SN}(p) &=& \frac{\Pi_{SN}}{(p^2-M_S^2)(p^2-M_N^2)-|\Pi_{SN}|^2}
\,.
\end{eqnarray}
\end{subequations}
It should be emphasized that, once the symmetry is broken dynamically by the above
defined anomalous propagators, nothing any longer prevents the
scalars from acquiring nonzero VEVs. In general, they will be induced by perturbative loop
effects. However, since it is our aim here to demonstrate the possibility of dynamical
EWSB, \emph{not driven} by scalar VEVs, we will neglect these VEVs altogether in our
Schwinger--Dyson (SD) equations. Their self-consistent incorporation will be studied
in a future work.

The self-energies $\Sigma$, $\Pi$ are subject to the SD equations,
which we approximate by the following set (see Fig.~\ref{fig_SD}):
\begin{subequations}
\label{eqn:SD}
\begin{eqnarray}
\Sigma_U &=&
\I y_U^2 \int\frac{\d^4k}{(2\pi)^4} D_N(k) S_U^*(k+p)
\\ && \nonumber
+ \I y_Uy_D \int\frac{\d^4k}{(2\pi)^4} D_{SN}(k) S_D^*(k+p)
\,,
\\
\Sigma_D &=&
\I y_D^2 \int\frac{\d^4k}{(2\pi)^4} D_S(k) S_D^*(k+p)
\\ && \nonumber
+ \I y_Uy_D \int\frac{\d^4k}{(2\pi)^4} D_{SN}(k) S_U^*(k+p)
\,,
\\
\Pi_S &=& -2\I\sum_D N_C y_D^{2} \int\frac{\d^4k}{(2\pi)^4} S_D(k) S_D(k+p)
\,,
\label{eqn:SD_pi_S}
\\
\Pi_N &=& -2\I\sum_U N_C y_U^{2} \int\frac{\d^4k}{(2\pi)^4} S_U(k) S_U(k+p)
\,,
\label{eqn:SD_pi_N}
\\
\Pi_{SN} &=& -2\I\sum_{\mathcal{D}} N_C y_Uy_D \int\frac{\d^4k}{(2\pi)^4} S_U(k)
S_D(k+p)
\,. \quad\quad
\label{eqn:SD_pi_SN}
\end{eqnarray}
\end{subequations}
($N_C$ denotes the number of colors). The sum in the last equation is over all
isospin doublets, $\mathcal{D}=(U,D)^T$. In our case, $\mathcal{D}=\ell,q$. The
numerical solution of these equations is discussed in Sec.~\ref{sec_results}.

\section{Gauge bosons}
\label{sec_gauged}
Since the $SU(2)_L \times U(1)_Y$ symmetry is broken down to
electromagnetic $U(1)_{em}$, the question about the nature of the corresponding NG
bosons (to become the longitudinal parts of the $W^\pm$, $Z$ bosons) arises. It turns out
that, according to general principles, they can be `seen' as massless poles in the proper
vertices of the currents associated with the broken generators \cite{Jackiw:1973tr}.
The residua of these poles are proportional to the symmetry-breaking proper self-energies
$\Sigma$ and $\Pi$.

Without going too much into technical details (interested reader can
find them in \cite{Benes:2006ny}), let us only sketch the way
leading to computation of the gauge boson masses. In this context it is useful
to consider the vertex functions of the broken currents,
\begin{subequations}
\begin{eqnarray}
G^\mu_{\mathcal{D} a}(x,y,z) & = & \bra{0}T\{j^\mu_a(x)\mathcal{D}(y)\mathcal{\bar
D}(z)\}\ket{0}
\,,
\\
G^\mu_{\Phi a}(x,y,z) & = & \bra{0}T\{j^\mu_a(x)\Phi(y)\Phi^{\dag}(z)\}\ket{0}
\,.
\end{eqnarray}
\end{subequations}
Here $j^\mu_a$ ($a=1,\ldots,4$) are the conserved currents associated
with the generators
of $SU(2)_L \times U(1)_Y$  and $\Phi=\Phi_S,\Phi_N,\Phi_{SN}$. Taking
the divergence of the vertex functions we
arrive at the Ward--Takahashi (WT) identities for the proper
vertices of the gauge bosons with the fermions and scalars. For instance,
for the proper vertex $Zf\bar f$ we get (similarly for other gauge bosons
and fermions/scalars)
\begin{equation}
q_\mu \Gamma^\mu_{Zf\bar f}(p^{\prime},p) =
\I\langle f \bar f \rangle^{-1}_{p^{\prime}} T_f -
\bar T_f \I\langle f \bar f \rangle^{-1}_p
\,,
\end{equation}
with $q=p^{\prime}-p$ denoting the momentum of the incoming gauge boson and
\begin{equation}
T_f = \frac{g}{2\cos\theta_W} \Big[ T_{3f}(1-\gamma_5) - 2Q_f\sin^2\theta_W \Big]
\,,
\end{equation}
and $\bar T_f=\gamma_0 T_f^\dag \gamma_0$. ($T_3$ and $Q$ denote the isospin and
charge matrices, diagonal in the flavor space.) These WT identities remain valid even
though the corresponding symmetries are spontaneously broken. Consequently, it is easy to see that upon
inserting the full propagators \eqref{ansatz_S}, \eqref{ansatz_D} in the WT identities and
taking the limit $q \rightarrow 0$, the proper vertices $\Gamma^\mu$ associated with the
broken symmetries (i.e., $\Gamma^\mu_{W^\pm}$, $\Gamma^\mu_{Z}$) develop
a massless pole. It is to be interpreted as the propagator of an intermediate
scalar excitation, bilinearly coupled to the gauge boson -- the NG boson. In the
case of $\Gamma^\mu_{Zf\bar f}$ its
pole part has the form ($\hat\Sigma_f \equiv
\Sigma_{f,1}+\I\gamma_5\Sigma_{f,2}$)
\begin{equation}
\Gamma^\mu_{Zf\bar f\mathrm{pole}}(p^{\prime},p) = \frac{q^\mu}{q^2} \left[
-\hat\Sigma_f(p^{\prime})T_f+\bar T_f\hat\Sigma_f(p) \right]
\,.
\end{equation}

\begin{figure}[t]
\begin{center}
\includegraphics[width=0.4\textwidth]{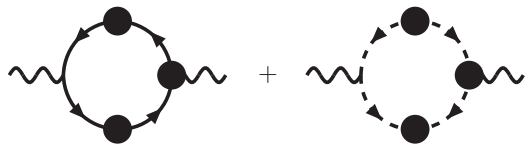}
\caption[]{Contributions to the polarization tensor of a gauge boson ($W^\pm$ or
$Z$) from the fermion and scalar loops. The first vertex in each loop is
bare, while the second one is the pole part of the proper
vertex, extracted from the WT identity.}
\label{fig_gauge}
\end{center}
\end{figure}

Since the broken symmetry is gauged, the NG bosons decouple from the spectrum and
manifest themselves only as the longitudinal polarizations of the $W^\pm$ and
$Z$ bosons. By the Schwinger mechanism \cite{Schwinger:1957em}, these poles in turn give
rise to the gauge boson masses, which are determined by the corresponding residues
\cite{Jackiw:1973tr}. Within our approximation, the polarization tensors are given by the
graphs in Fig.~\ref{fig_gauge} with the insertion of the pole parts of the proper
vertices.

As a net result, the gauge boson masses can be expressed via the sum rules
\begin{subequations}
\label{eqn:sum_rules}
\begin{eqnarray}
M_Z^2 & = & \frac{g^2}{\cos^2\theta_W} \Big( I_{S}+I_{N}+\sum_f I_{f} \Big)
\,,
\\
M_W^2 & = & g^2 \Big( I_{S,SN}+I_{N,SN}+\sum_{\mathcal{D}} I_{\mathcal{D}}\Big)
\,,
\end{eqnarray}
\end{subequations}
with
\begin{subequations}
\label{eqn:I}
\begin{eqnarray}
I_{X} & = & \I\frac{1}{2} \int\frac{\d^4 k}{(2\pi)^4} k^2 |D_X(k)|^2
\,,
\\
I_{X,SN} & = & \I\frac{1}{4} \int\frac{\d^4 k}{(2\pi)^4}
k^2 D_X(k)D_{SN}(k) \frac{|\Pi_X|^2+|\Pi_{SN}|^2}{\Pi_X\Pi_{SN}}
\,,
\nonumber \\ &&
\\
I_{f} & = & -\I\frac{1}{2} N_C \int\frac{\d^4 k}{(2\pi)^4} |S_f(k)|^2
\,,
\\
I_{\mathcal{D}} & = & -\I\frac{1}{2} N_C \int\frac{\d^4 k}{(2\pi)^4}
S_U(k)S_D(k) \frac{|\Sigma_U|^2+|\Sigma_D|^2}{\Sigma_U\Sigma_D}
\,.
\nonumber \\ &&
\end{eqnarray}
\end{subequations}
Thanks to the fact that the dynamically generated self-energies are suppressed at
high momentum, all integrals above are UV-finite.

\section{Results and discussion}
\label{sec_results}

The numerical analysis of the (Wick-rotated) SD equations \eqref{eqn:SD}
shows that: (i) Non-trivial, UV-finite solutions do exist. (ii) The solutions
are found only for relatively large values of the Yukawa coupling constants (of
order of tens). (iii) Large ratios of fermionic masses can be accommodated while
having the corresponding Yukawa coupling constants of the same order of
magnitude.

The point (iii) above is promising in the
quest for realistic fermion mass hierarchy. Because of the large parameter
space which needs to be scanned this has not been accomplished yet.
However, some achievements, which suggest that it should be possible,
have been made. First, we accommodated the hierarchy between the lepton
and quark doublets. For $y_\nu=63$, $y_e\approx84$, $y_u=65$, $y_d=90$ (and $M_S^2=2$,
$M_N^2=1$) we found $m_\nu \apprle m_e=\mathcal{O}(10^{-4})$ and $m_u  \apprle
m_d=\mathcal{O}(10^{-2})$. (Note that all masses are expressed in the units of
$M_N$.) Second, we managed to generate a large hierarchy within one doublet. Considering
only the leptons and `neglecting' the quarks ($y_u=y_d=0$), we found
$m_e/m_\nu=\mathcal{O}(10^{2})$, calculated for $y_\nu\approx50$, $y_e=80$ (and again
$M_S^2=2$, $M_N^2=1$). Nevertheless, it should be emphasized that this lepton
mass ratio will be significantly enhanced by the seesaw mechanism once
the Majorana mass term is taken into account.

The numerical results above were
based on the exact solution of the Wick-rotated SD equations. In
this numerical procedure one chooses some values of the input parameters
(the Yukawa coupling constants and the bare scalar masses), solves the equations
and checks whether the resulting spectrum (fermion and gauge boson masses)
corresponds to the real values. It is, however, difficult to
find in this manner the combination of the parameters, giving realistic
spectrum; not only because of the hugeness of the parameter space, but also
since solving the SD equations even for a single set of parameters is
rather time-consuming.

Thus, in order to be able to invert the above procedure, i.e., to choose the
spectrum first and then find the corresponding parameters of the model, we
relaxed the exactness of the solution and solved the SD equations only
approximately. For that purpose we chose the Ansatz for the fermion
and scalar self-energies in the form of step-functions with a
common position of the step $\Lambda$ (i.e., $\Sigma_f(p^2)\rightarrow
m_f\heaviside(\Lambda^2-p^2)$
and $\Pi_X(p^2)\rightarrow\mu_X\heaviside(\Lambda^2-p^2)$)
and plugged them into the SD equations. This Ansatz is decently justified
by the observation that the self-energies obtained by exact solution
of the SD equations exhibit similar behavior: They are nearly constant in a
wide range of momenta and then drop fast to zero around a scale common to all
particles in the model.

With an Ansatz on the shape of the self-energies, the SD equations
can of course no longer be satisfied identically. Because mass
generation is a low-energy phenomenon, we demand instead that the left-
and right-hand sides are equal at $p^2=0$. This results in a set of several
algebraic equations for the unknown parameters of the model, which can be solved
much more easily than the full integral equations \eqref{eqn:SD}.

According to the sum rules \eqref{eqn:sum_rules}, the heavy quark flavors
contribute most to the gauge boson masses. Likewise, they also dominate the scalar self-energies \eqref{eqn:SD_pi_S}--\eqref{eqn:SD_pi_SN}, which tie the individual fermion masses together. We therefore neglected for the moment
the lepton doublet and treated the quark one as the $t$- and $b$-quarks. We
kept the gauge boson masses at the physical values. Due to
numerical reasons, however, we could only fix the quark masses to the rather
large values $m_t=405\,\mathrm{GeV}$,
$m_b=10\,\mathrm{GeV}$. Nevertheless, note that the ratio of
these masses is still realistic. We then found the following solution:
\begin{subequations}
\begin{eqnarray}
y_t &=& 375 \,, \\
y_b &=& 235 \,, \\
M_N &=& 433\,\mathrm{TeV} \,, \\
M_S &=& 283\,\mathrm{TeV} \,, \\
\Lambda &=& 71\,\mathrm{TeV} \,.
\end{eqnarray}
\end{subequations}
This result shows again some important features. First, the Yukawa couplings are
of the same order of magnitude, while keeping the realistic ratio of the quark
masses in the doublet. Second, the scalar masses are much higher than the EW
scale, which is gratifying, as explained above. Our numerical investigations
suggest that if we managed to pull the quark masses down to the physical values,
the scalar masses would even increase by a few orders of magnitude.

The `cut-off' $\Lambda$ also deserves a remark.
It corresponds to the energy scale at which the
self-energies fall rapidly to zero. The fact that
$\Lambda \ll M_S, M_N$ suggests that the scalar self-energies are practically
vanishing for momenta corresponding to the scalar pole masses. Consequently, the
scalar mass splitting is negligible and the $S$-parameter tends to be
very close to zero. (For $\Pi_S=\Pi_N=\Pi_{SN}=0$ we would have exactly
$S=0$.) On the other hand, since $\Lambda \gg m_t, m_b$, the quark self-energies
are almost constant (and non-zero) at the energy scale of fermion masses (under
$1 \, \mathrm{TeV}$). If we considered all three fermion generations together with a
mixing, this would consequently lead to the CKM matrix being significantly close
to the unitary form \cite{Benes:2009iz}.

\subsection{Compatibility with electroweak observables}
While the realistic fermion spectrum together with the Yukawa coupling
constants can be presumably accommodated, it on the other hand brings the
problem how to keep the $\rho$-parameter (defined as $\rho\equiv
M_W^2/(M_Z^2\cos^2\theta_W)$) close to $1$. Note that if we set $y_\nu=y_e$,
$y_u=y_d$ and $M_S=M_N$, the SD equations \eqref{eqn:SD} yield
$\Sigma_\nu=\Sigma_e$, $\Sigma_u=\Sigma_d$ and $\Pi_S=\Pi_N=\Pi_{SN}$ and
consequently $\rho=1$ exactly. (Which is not surprising,
since in this case the Lagrangian possesses the custodial symmetry.) Of course,
in reality  $y_\nu\neq y_e$, $y_u \neq y_d$, so the custodial symmetry is
broken by the fermions. However, for the values $m_\nu \apprle
m_e=\mathcal{O}(10^{-4})$ and $m_u \apprle m_d=\mathcal{O}(10^{-2})$ mentioned
above, we find $\rho \doteq 1.012$. It is thus apparent that one can achieve
realistic fermion mass ratios and still keep $\rho$ reasonably close to one. A
detailed analysis reveals that key r\^{o}le is here played by the scalars: If
their bare masses and the self-energies are of the same order of magnitude
(i.e., $M_S \approx M_N$ and $\Pi_S \approx \Pi_N \approx \Pi_{SN}$) so that
they approximately conserve the custodial symmetry, they can render $\rho$
close to $1$, provided they are heavy enough so that
they can overcome the effect of the custodial symmetry breaking in the fermion
sector.

The fact that the scalars tend to be heavy
is in fact reassuring. First, we do not have to deal with the usual hierarchy
problem, that is, radiative instability of the masses of elementary scalars at
the TeV scale. Of course, the problem is just postponed to a higher scale. In
this sense we understand our model as an effective approach which is valid at
energy scales around and below the heavy scalar masses. We do not pretend to
have an ultraviolet complete theory.

Second, the new
scalars must be heavy enough in order to avoid constraints from flavor-changing
neutral currents (FCNC). Even though there are apparently no FCNC in
the present model, for we only consider a single family at the moment,
anticipating a future extension to all three families of SM fermions, we can
make at least a rough, order-of-magnitude estimate. With several fermion
families there will be explicit flavor-changing Yukawa interactions, allowing,
for instance, for the decay $\mu^-\to e^-S^{(0)}$. The virtual heavy scalar can
subsequently decay as $S^{(0)}\to e^+e^-$. Our Yukawa interactions will
therefore induce the flavor-chaning muon decay, $\mu^-\to e^-e^+e^-$, with the
amplitude being roughly given by $y^2/M_S^2$. (We assume that in the absence of
fine tuning, all Yukawa couplings, including the flavor-changing ones, will be
of the same order of magnitude.) The dominant muon decay channel, with branching
ratio close to $100\%$, is $\mu^-\to e^-\bar\nu_e\nu_\mu$, whose amplitude is
analogously proportional to $G_F$. From here we infer the estimate
$\BR(\mu^-\to e^-e^+e^-)\sim(y^2/G_FM_S^2)^2$. Taking the current
experimental limit \cite{Amsler:2008zzb}, $\BR(\mu^-\to e^-e^+e^-)<10^{-12}$, we
find $M_S/y\apprge10^{2.5}\,\mathrm{TeV}$. Our numerical
calculations reported above suggest that this constraint might impose some
tension on our model, but it is certainly possible to satisfy.

\begin{figure}[t]
\begin{center}
\framebox{\scalebox{0.8}{\include{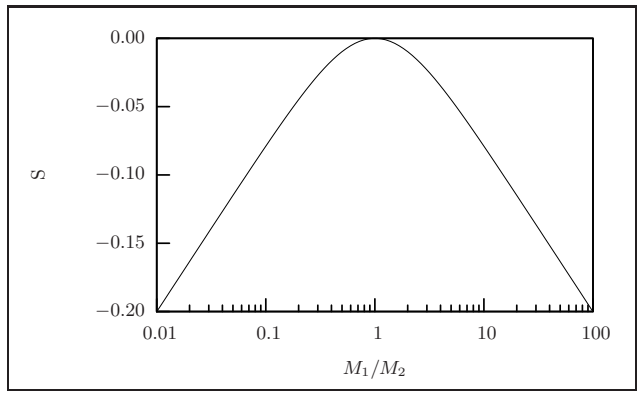}}}
\end{center}
\caption{The $S$-parameter plotted for the special case $M_{S1,2} = M_{N1,2} =
M_{SN1,2} \equiv M_{1,2}$. Note that, according to the Particle Data Group
\cite{Amsler:2008zzb}, $S = -0.10 \pm 0.10 $.}
\label{fig_plot_S}
\end{figure}

The very introduction of new scalars also affects the Peskin--Takeuchi
$S$-parameter \cite{Peskin:1991sw}. In order to estimate the scalar contribution to it, we
set for simplicity the scalar self-energies to be constant. The resulting
$S$-parameter can then be written as $S=S_{S} + S_{N} + S_{SN}$, where
\begin{subequations}
\begin{eqnarray}
S_{X} & = &  \frac{1}{12 \pi}
\Bigg[
\frac{5}{6} - \frac{2M_{X1}^2 M_{X2}^2}{\big(M_{X1}^2-M_{X2}^2\big)^2}
-\frac{1}{2}\ln\frac{M_{X1}^2 M_{X2}^2}{\mu^4}
\nonumber \\ && \!\!\!\!\!\!\!\!\!\!\!\!\!\!\!\!\!\!\!\!\!\!
-\frac{1}{2} \frac{M_{X1}^6+M_{X2}^6 - 3 M_{X1}^2M_{X2}^2
(M_{X1}^2+M_{X2}^2)}{\big(M_{X1}^2-M_{X2}^2\big)^3}\ln\frac{M_{X1}^2}{M_{X2}^2}
\Bigg] \,, \nonumber
\\ &&
\\
S_{SN} & = &  \frac{1}{12 \pi}
\ln \frac{M_{SN1}^2 M_{SN2}^2}{\mu^4} \,.
\end{eqnarray}
\end{subequations}
(The $\mu$ is just an arbitrary mass scale introduced for convenience; the total
$S$-parameter is independent of it.) Taking into account the previous discussion
of the $\rho$-parameter and the scalar masses, we plotted the $S$-parameter for
the special case $M_{S1,2} = M_{N1,2} = M_{SN1,2} \equiv M_{1,2}$. The resulting
$S$-parameter (see Fig.~\ref{fig_plot_S}) meets the experimental bounds for any
value of the mass ratio $M_1/M_2$ from $0.01$ up to $100$. When in this
special case the ratio $M_1/M_2$ is far from one, the $S$-parameter is well
approximated by the simple formula
$S=\frac{1}{6\pi}\bigl(\frac56-\ln\bigl|\frac{M_1}{M_2}\bigr|\bigr)$.
On the other hand, for $M_1/M_2$ close to one the $S$-parameter behaves like $\frac{-1}{15 \pi}\bigl(1-\frac{M_1}{M_2}\bigr)^2$.

\section{Conclusions and outlook}
We have presented a new possible mechanism and the corresponding formalism for
EWSB. We have introduced elementary scalars and argued that EWSB could be driven
not by their VEVs, like in SM, but rather dynamically, by their Yukawa couplings
to fermions. If this was the case, then the `would-be' NG bosons should be, as
shown by general arguments, composites of both the fermions and the scalars.

The presented model certainly has some advantages: (i) It distinguishes the
fermion species already at the level of Lagrangian by their distinct Yukawa
couplings and hence avoids the dangerous flavor symmetries. (ii) It ties the
gauge boson masses up with the fermion and scalar masses by means of the sum
rules \eqref{eqn:sum_rules}. (iii) Since the resulting fermion masses are
nonlinear functions of the Yukawa coupling constants, there is a possibility to
accommodate large ratios of fermion masses, as seen in experiment, while keeping
the Yukawa couplings of the same order of magnitude. This is supported by the
numerical results shown above, as well as by the numerical analysis of our
preceding model \cite{Benes:2006ny}, dealing with a simpler toy version of the
present model of strong Yukawa dynamics.

So far we have considered only one fermionic generation. Including all three
generations will require to introduce the flavor mixing through the Yukawa
coupling constant matrices. This opens a new way to study dynamical CP violation in both
the quark and lepton sector, since even though the Yukawa coupling constants are real, the
solutions to the SD equations are in general complex \cite{Bicudo:2003fd},
hence leading to complex `mixing matrices'.

Upon necessary modifications, the present mechanism has also the potential to
describe non-relativistic bosonic superfluidity without a single-field
condensate. The idea is supported by Refs.~\cite{Imry} and
\cite{PhysRevLett.89.075301}, in which the Cooper-type pairing in the many-boson
system of~$^{4}\mathrm{He}$ is studied. The anomalous Green's function of the
type $\langle\Phi\Phi\rangle$, considered here, was also studied in Ref.~\cite{Alexandrov2004}, along with the conventional Bose condensate.

\begin{acknowledgments}
The authors gratefully acknowledge discussions with J.~Ho\v{s}ek and
H.~B\'{\i}la. This work was supported in part by the Institutional Research
Plan AV0Z10480505, and by the GACR Grant No.~202/06/0734. The work of T.~B.~was
also supported by the Alexander von Humboldt Foundation.
\end{acknowledgments}


\begin{thebibliography}{29}
\expandafter\ifx\csname natexlab\endcsname\relax\def\natexlab#1{#1}\fi
\expandafter\ifx\csname bibnamefont\endcsname\relax
  \def\bibnamefont#1{#1}\fi
\expandafter\ifx\csname bibfnamefont\endcsname\relax
  \def\bibfnamefont#1{#1}\fi
\expandafter\ifx\csname citenamefont\endcsname\relax
  \def\citenamefont#1{#1}\fi
\expandafter\ifx\csname url\endcsname\relax
  \def\url#1{\texttt{#1}}\fi
\expandafter\ifx\csname urlprefix\endcsname\relax\def\urlprefix{URL }\fi
\providecommand{\bibinfo}[2]{#2}
\providecommand{\eprint}[2][]{\url{#2}}

\bibitem[{\citenamefont{Weinberg}(1967)}]{Weinberg:1967tq}
\bibinfo{author}{\bibfnamefont{S.}~\bibnamefont{Weinberg}},
  \bibinfo{journal}{Phys. Rev. Lett.} \textbf{\bibinfo{volume}{19}},
  \bibinfo{pages}{1264} (\bibinfo{year}{1967}).

\bibitem[{\citenamefont{Lee}(1973)}]{Lee:1973iz}
\bibinfo{author}{\bibfnamefont{T.~D.} \bibnamefont{Lee}},
  \bibinfo{journal}{Phys. Rev.} \textbf{\bibinfo{volume}{D8}},
  \bibinfo{pages}{1226} (\bibinfo{year}{1973}).

\bibitem[{\citenamefont{Haber and Kane}(1985)}]{Haber:1984rc}
\bibinfo{author}{\bibfnamefont{H.~E.} \bibnamefont{Haber}} \bibnamefont{and}
  \bibinfo{author}{\bibfnamefont{G.~L.} \bibnamefont{Kane}},
  \bibinfo{journal}{Phys. Rept.} \textbf{\bibinfo{volume}{117}},
  \bibinfo{pages}{75} (\bibinfo{year}{1985}).

\bibitem[{\citenamefont{Arkani-Hamed et~al.}(2001)\citenamefont{Arkani-Hamed,
  Cohen, and Georgi}}]{ArkaniHamed:2001nc}
\bibinfo{author}{\bibfnamefont{N.}~\bibnamefont{Arkani-Hamed}},
  \bibinfo{author}{\bibfnamefont{A.~G.} \bibnamefont{Cohen}}, \bibnamefont{and}
  \bibinfo{author}{\bibfnamefont{H.}~\bibnamefont{Georgi}},
  \bibinfo{journal}{Phys. Lett.} \textbf{\bibinfo{volume}{B513}},
  \bibinfo{pages}{232} (\bibinfo{year}{2001}), \eprint{hep-ph/0105239}.

\bibitem[{\citenamefont{Arkani-Hamed et~al.}(2002)}]{Arkani-Hamed:2002qx}
\bibinfo{author}{\bibfnamefont{N.}~\bibnamefont{Arkani-Hamed}}
  \bibnamefont{et~al.}, \bibinfo{journal}{JHEP} \textbf{\bibinfo{volume}{08}},
  \bibinfo{pages}{021} (\bibinfo{year}{2002}), \eprint{hep-ph/0206020}.

\bibitem[{\citenamefont{Weinberg}(1979)}]{Weinberg:1979bn}
\bibinfo{author}{\bibfnamefont{S.}~\bibnamefont{Weinberg}},
  \bibinfo{journal}{Phys. Rev.} \textbf{\bibinfo{volume}{D19}},
  \bibinfo{pages}{1277} (\bibinfo{year}{1979}).

\bibitem[{\citenamefont{Susskind}(1979)}]{Susskind:1978ms}
\bibinfo{author}{\bibfnamefont{L.}~\bibnamefont{Susskind}},
  \bibinfo{journal}{Phys. Rev.} \textbf{\bibinfo{volume}{D20}},
  \bibinfo{pages}{2619} (\bibinfo{year}{1979}).

\bibitem[{\citenamefont{Dimopoulos and Susskind}(1979)}]{Dimopoulos:1979es}
\bibinfo{author}{\bibfnamefont{S.}~\bibnamefont{Dimopoulos}} \bibnamefont{and}
  \bibinfo{author}{\bibfnamefont{L.}~\bibnamefont{Susskind}},
  \bibinfo{journal}{Nucl. Phys.} \textbf{\bibinfo{volume}{B155}},
  \bibinfo{pages}{237} (\bibinfo{year}{1979}).

\bibitem[{\citenamefont{Eichten and Lane}(1980)}]{Eichten:1979ah}
\bibinfo{author}{\bibfnamefont{E.}~\bibnamefont{Eichten}} \bibnamefont{and}
  \bibinfo{author}{\bibfnamefont{K.~D.} \bibnamefont{Lane}},
  \bibinfo{journal}{Phys. Lett.} \textbf{\bibinfo{volume}{B90}},
  \bibinfo{pages}{125} (\bibinfo{year}{1980}).

\bibitem[{\citenamefont{Miransky et~al.}(1989)\citenamefont{Miransky,
  Tanabashi, and Yamawaki}}]{Miransky:1988xi}
\bibinfo{author}{\bibfnamefont{V.~A.} \bibnamefont{Miransky}},
  \bibinfo{author}{\bibfnamefont{M.}~\bibnamefont{Tanabashi}},
  \bibnamefont{and} \bibinfo{author}{\bibfnamefont{K.}~\bibnamefont{Yamawaki}},
  \bibinfo{journal}{Phys. Lett.} \textbf{\bibinfo{volume}{B221}},
  \bibinfo{pages}{177} (\bibinfo{year}{1989}).

\bibitem[{\citenamefont{Bardeen et~al.}(1990)\citenamefont{Bardeen, Hill, and
  Lindner}}]{Bardeen:1989ds}
\bibinfo{author}{\bibfnamefont{W.~A.} \bibnamefont{Bardeen}},
  \bibinfo{author}{\bibfnamefont{C.~T.} \bibnamefont{Hill}}, \bibnamefont{and}
  \bibinfo{author}{\bibfnamefont{M.}~\bibnamefont{Lindner}},
  \bibinfo{journal}{Phys. Rev.} \textbf{\bibinfo{volume}{D41}},
  \bibinfo{pages}{1647} (\bibinfo{year}{1990}).

\bibitem[{\citenamefont{Hill}(1991)}]{Hill:1991at}
\bibinfo{author}{\bibfnamefont{C.~T.} \bibnamefont{Hill}},
  \bibinfo{journal}{Phys. Lett.} \textbf{\bibinfo{volume}{B266}},
  \bibinfo{pages}{419} (\bibinfo{year}{1991}).

\bibitem[{\citenamefont{Hill}(1995)}]{Hill:1994hp}
\bibinfo{author}{\bibfnamefont{C.~T.} \bibnamefont{Hill}},
  \bibinfo{journal}{Phys. Lett.} \textbf{\bibinfo{volume}{B345}},
  \bibinfo{pages}{483} (\bibinfo{year}{1995}), \eprint{hep-ph/9411426}.

\bibitem[{\citenamefont{Nambu and Jona-Lasinio}(1961)}]{Nambu:1961tp}
\bibinfo{author}{\bibfnamefont{Y.}~\bibnamefont{Nambu}} \bibnamefont{and}
  \bibinfo{author}{\bibfnamefont{G.}~\bibnamefont{Jona-Lasinio}},
  \bibinfo{journal}{Phys. Rev.} \textbf{\bibinfo{volume}{122}},
  \bibinfo{pages}{345} (\bibinfo{year}{1961}).

\bibitem[{\citenamefont{{Ho\v{s}ek}}(1982)}]{Hosek:1982cz}
\bibinfo{author}{\bibfnamefont{J.}~\bibnamefont{{Ho\v{s}ek}}}
  (\bibinfo{year}{1982}), \bibinfo{note}{submitted to 21st Int. Conf. on High
  Energy Physics, Paris, France, Jul 26-31, 1982}.

\bibitem[{\citenamefont{Wetterich}(2006)}]{Wetterich:2006ii}
\bibinfo{author}{\bibfnamefont{C.}~\bibnamefont{Wetterich}},
  \bibinfo{journal}{Phys. Rev.} \textbf{\bibinfo{volume}{D74}},
  \bibinfo{pages}{095009} (\bibinfo{year}{2006}), \eprint{hep-ph/0607051}.

\bibitem[{\citenamefont{Luty et~al.}(2001)\citenamefont{Luty, Terning, and
  Grant}}]{Luty:2000fj}
\bibinfo{author}{\bibfnamefont{M.~A.} \bibnamefont{Luty}},
  \bibinfo{author}{\bibfnamefont{J.}~\bibnamefont{Terning}}, \bibnamefont{and}
  \bibinfo{author}{\bibfnamefont{A.~K.} \bibnamefont{Grant}},
  \bibinfo{journal}{Phys. Rev.} \textbf{\bibinfo{volume}{D63}},
  \bibinfo{pages}{075001} (\bibinfo{year}{2001}), \eprint{hep-ph/0006224}.

\bibitem[{\citenamefont{Imry}(1970)}]{Imry}
\bibinfo{author}{\bibfnamefont{Y.}~\bibnamefont{Imry}},
  \emph{\bibinfo{title}{Self consistent pairing theory of the Bose
  superfluids}} (\bibinfo{publisher}{Gordon \& Breach, New York},
  \bibinfo{year}{1970}), pp. \bibinfo{pages}{603--614}, \bibinfo{note}{in
  Quantum Fluids, Proceedings of the Batsheva Seminar, Haifa, 1968, eds.: N.
  Wiser and D. J. Amit}.

\bibitem[{\citenamefont{Pashitskii et~al.}(2002)\citenamefont{Pashitskii,
  Mashkevich, and Vilchynskyy}}]{PhysRevLett.89.075301}
\bibinfo{author}{\bibfnamefont{E.~A.} \bibnamefont{Pashitskii}},
  \bibinfo{author}{\bibfnamefont{S.~V.} \bibnamefont{Mashkevich}},
  \bibnamefont{and} \bibinfo{author}{\bibfnamefont{S.~I.}
  \bibnamefont{Vilchynskyy}}, \bibinfo{journal}{Phys. Rev. Lett.}
  \textbf{\bibinfo{volume}{89}}, \bibinfo{pages}{075301}
  (\bibinfo{year}{2002}).

\bibitem[{\citenamefont{Alexandrov}(2004)}]{Alexandrov2004}
\bibinfo{author}{\bibfnamefont{A.~S.} \bibnamefont{Alexandrov}}
  (\bibinfo{year}{2004}), \eprint{arXiv:cond-mat/0402123}.

\bibitem[{\citenamefont{{Bene\v{s}} et~al.}(2007)\citenamefont{{Bene\v{s}},
  Brauner, and {Ho\v{s}ek}}}]{Benes:2006ny}
\bibinfo{author}{\bibfnamefont{P.}~\bibnamefont{{Bene\v{s}}}},
  \bibinfo{author}{\bibfnamefont{T.}~\bibnamefont{Brauner}}, \bibnamefont{and}
  \bibinfo{author}{\bibfnamefont{J.}~\bibnamefont{{Ho\v{s}ek}}},
  \bibinfo{journal}{Phys. Rev.} \textbf{\bibinfo{volume}{D75}},
  \bibinfo{pages}{056003} (\bibinfo{year}{2007}), \eprint{hep-ph/0605147}.

\bibitem[{\citenamefont{Bornholdt and Wetterich}(1993)}]{Bornholdt:1992za}
\bibinfo{author}{\bibfnamefont{S.}~\bibnamefont{Bornholdt}} \bibnamefont{and}
  \bibinfo{author}{\bibfnamefont{C.}~\bibnamefont{Wetterich}},
  \bibinfo{journal}{Z. Phys.} \textbf{\bibinfo{volume}{C58}},
  \bibinfo{pages}{585} (\bibinfo{year}{1993}).

\bibitem[{\citenamefont{Gies et~al.}(2004)\citenamefont{Gies, Jaeckel, and
  Wetterich}}]{Gies:2003dp}
\bibinfo{author}{\bibfnamefont{H.}~\bibnamefont{Gies}},
  \bibinfo{author}{\bibfnamefont{J.}~\bibnamefont{Jaeckel}}, \bibnamefont{and}
  \bibinfo{author}{\bibfnamefont{C.}~\bibnamefont{Wetterich}},
  \bibinfo{journal}{Phys. Rev.} \textbf{\bibinfo{volume}{D69}},
  \bibinfo{pages}{105008} (\bibinfo{year}{2004}), \eprint{hep-ph/0312034}.

\bibitem[{\citenamefont{Jackiw and Johnson}(1973)}]{Jackiw:1973tr}
\bibinfo{author}{\bibfnamefont{R.}~\bibnamefont{Jackiw}} \bibnamefont{and}
  \bibinfo{author}{\bibfnamefont{K.}~\bibnamefont{Johnson}},
  \bibinfo{journal}{Phys. Rev.} \textbf{\bibinfo{volume}{D8}},
  \bibinfo{pages}{2386} (\bibinfo{year}{1973}).

\bibitem[{\citenamefont{Schwinger}(1957)}]{Schwinger:1957em}
\bibinfo{author}{\bibfnamefont{J.~S.} \bibnamefont{Schwinger}},
  \bibinfo{journal}{Annals Phys.} \textbf{\bibinfo{volume}{2}},
  \bibinfo{pages}{407} (\bibinfo{year}{1957}).

\bibitem[{\citenamefont{Bene\v{s}}(2009)}]{Benes:2009iz}
\bibinfo{author}{\bibfnamefont{P.}~\bibnamefont{Bene\v{s}}}
  (\bibinfo{year}{2009}), \eprint{arXiv:0904.0139 [hep-ph]}.

\bibitem[{\citenamefont{Amsler et~al.}(2008)}]{Amsler:2008zzb}
\bibinfo{author}{\bibfnamefont{C.}~\bibnamefont{Amsler}} \bibnamefont{et~al.}
  (\bibinfo{collaboration}{Particle Data Group}), \bibinfo{journal}{Phys.
  Lett.} \textbf{\bibinfo{volume}{B667}}, \bibinfo{pages}{1}
  (\bibinfo{year}{2008}).

\bibitem[{\citenamefont{Peskin and Takeuchi}(1992)}]{Peskin:1991sw}
\bibinfo{author}{\bibfnamefont{M.~E.} \bibnamefont{Peskin}} \bibnamefont{and}
  \bibinfo{author}{\bibfnamefont{T.}~\bibnamefont{Takeuchi}},
  \bibinfo{journal}{Phys. Rev.} \textbf{\bibinfo{volume}{D46}},
  \bibinfo{pages}{381} (\bibinfo{year}{1992}).

\bibitem[{\citenamefont{Bicudo}(2004)}]{Bicudo:2003fd}
\bibinfo{author}{\bibfnamefont{P.}~\bibnamefont{Bicudo}},
  \bibinfo{journal}{Phys. Rev.} \textbf{\bibinfo{volume}{D69}},
  \bibinfo{pages}{074003} (\bibinfo{year}{2004}), \eprint{hep-ph/0312373}.

\end{thebibliography}

\end{document}